\documentclass[iop,apj,onecolumn,tighten]{emulateapj}
\usepackage{mathrsfs}
\begin{document}

\title{Numerical Simulations of Optically Thick Accretion onto a Black Hole - I. Spherical Case}
\shorttitle{Optically Thick Spherical Accretion}
\shortauthors{Fragile, et al.}
\author{P. Chris Fragile\footnotemark[1], Anna Gillespie, Timothy Monahan, and Marco Rodriguez}
\affil{Department of Physics \& Astronomy, College of Charleston, Charleston, SC 29424, USA}
\and
\author{Peter Anninos}
\affil{Lawrence Livermore National Laboratory, P.O. Box 808, Livermore, CA 94550, USA}

\begin{abstract}
Modeling the radiation generated by accreting matter is an important step towards realistic simulations of black hole accretion disks, especially at high accretion rates. To this end, we have recently added radiation transport to the existing general relativistic magnetohydrodynamic code, {\em Cosmos++}.  However, before attempting to model radiative accretion disks, we have tested the new code using a series of shock tube and Bondi (spherical inflow) problems.  The four radiative shock tube tests, first presented by Farris et al. (2008), have known analytic solutions, allowing us to calculate errors and convergence rates for our code.  The Bondi problem only has an analytic solution when radiative processes are ignored, but is pertinent because it is closer to the physics we ultimately want to study.  In our simulations, we include Thomson scattering and thermal bremsstrahlung in the opacity, focusing exclusively on the super-Eddington regime.  Unlike accretion onto bodies with solid surfaces, super-Eddington accretion onto black holes does not produce super-Eddington luminosity.  In our examples, despite accreting at up to 300 times the Eddington rate, our measured luminosity is always several orders of magnitude below Eddington.  
\end{abstract}

\footnotetext[1]{KITP Visiting Scholar, Kavli Institute for Theoretical Physics, Santa Barbara, CA.}

\section{Introduction}
\label{sec:intro}

After more than a decade of successes in the area of general relativistic magnetohydrodynamic (GRMHD) numerical simulations of black hole accretion flows \citep[see][for a review]{abramowicz11}, there is now considerable understanding of accretion in this context.  However, one context in which our understanding is still quite limited, is the inner parts of luminous accretion flows.  Here radiation pressure is expected to dominate over gas pressure in supporting the flow against the gravitational field of the black hole.  Because of photon diffusion, such radiation-pressure-dominated regions can be even more fragile to the development of inhomogeneous density structures than gas-pressure-dominated ones, and because of the proximity to the black hole, relativistic effects are of utmost importance.  For these reasons, numerical study of this and similar problems demands the development of general relativistic {\em radiation} MHD codes.  

Although radiation hydrodynamics codes have been around for more than 4 decades \citep[e.g.][]{christy66,colgate66,cox66}, very few multidimensional, fully (general) relativistic treatments exist even today.  A few noteworthy examples closely related to our work were presented by \citet{devilliers08,farris08,zanotti11}.  Parallel progress has also been made in developing radiation hydrodynamics codes to solve the problem of core-collapse supernovae \citep{muller10,shibata11,lentz12}.  There the radiation is in the form of neutrinos, but the processes (and code challenges) are very similar.  The paucity of codes can be understood when one realizes that most radiation schemes become quite difficult to implement in multidimensions, given the large number of phase space degrees of freedom that need to be tracked for the radiation field.  The sheer computational demand of multidimensional radiation MHD has, until recently, been beyond the reach of even the largest scientific computers.  

The potential for discovery from such simulations, though, is significant.  The addition of radiative phenomena into general relativistic fluid dynamic simulations greatly expands the possible parameter space that can be explored, with interesting phenomena such as radiation-pressure-dominated slim disks \citep{abramowicz88}, secular instability \citep{lightman74}, photon bubbles \citep{begelman01}, and radiation driven outflows, all topics that could potentially be investigated.  Therefore, in this paper, we take the first, simplest step, considering spherical accretion onto a non-rotating black hole.  This case has been considered many times previously \citep{shapiro73,schmidburgk78,meier79,soffel82,vitello84,zampieri96}, but never, to our knowledge, with an explicit, multi-dimensional, Eulerian, general relativistic radiation MHD code such as the one implemented here.  For this reason, we feel it is worthwhile for us to present these results.

Although our implementation of GR radiation MHD is not unique -- in fact, it is very similar to at least two previous implementations described in the literature \citep{farris08,zanotti11} -- we, nevertheless, begin our paper in \S \ref{sec:method} with a detailed discussion of our method.  In \S \ref{sec:tests} we report on a series of one-dimensional shock tube tests meant to validate our code.  In \S \ref{sec:bondi} we arrive at the main results of this paper - one-dimensional spherical accretion onto a black hole {\em including radiation}.  We conclude in \S \ref{sec:conclusion}.  Most of the equations in this work are written in units where $GM=c=1$, although in a few places we leave in factors of $c$ for clarity.


\section{Numerical Method}
\label{sec:method}

\subsection{GR MHD}
We begin with a review of general relativistic MHD before discussing the addition of radiation.  This will make it easier for us to point out the advantages of our implementation.  
We start with the three basic laws of GRMHD:  the conservation of the stress-energy tensor
\begin{equation}
\left. T^{\alpha \beta}\right._{;\beta} = 0 ~,
\end{equation}
the continuity equation
\begin{equation}
\left(\rho u^\beta\right)_{;\beta} = 0 ~,
\end{equation}
and the homogenous Maxwell's equation 
\begin{equation}
\left. ^*F^{\alpha \beta}\right._{;\beta} = 0 ~.
\end{equation}
In the MHD limit, the stress-energy tensor can be expressed as
\begin{equation}
T^{\alpha \beta} = (\rho h + 2 P_\mathrm{mag}) u^\alpha u^\beta + (P_\mathrm{gas} + P_\mathrm{mag}) g^{\alpha \beta} - b^\alpha b^\beta
\end{equation}
where $\rho$ is the rest mass density, $h = 1 + \epsilon + P_\mathrm{gas}/\rho$ is the specific enthalpy, $\epsilon$ is the specific internal energy density, $P_\mathrm{gas}$ is the gas pressure, $P_\mathrm{mag}$ is the magnetic pressure, $u^\alpha = g^{\alpha \beta} u_\beta$ is the fluid 4-velocity, $g_{\alpha\beta}$ is the 4-metric, $g$ is the 4-metric determinant, $F^{\alpha \beta}$ is the Faraday tensor, and $^*F^{\alpha \beta} = \epsilon^{\alpha \beta \gamma \delta} F_{\gamma \delta}/2$ is its dual.  For this work we use a $\Gamma$-law equation of state (EOS),
\begin{equation}
P = (\Gamma-1)\rho\epsilon
\end{equation}
where $\Gamma$ (without subscripts or superscripts) is the adiabatic index.

Expanding the MHD equations and reorganizing in terms of new variables, we arrive at the following set of coupled equations
\begin{eqnarray}
 \partial_t D + \partial_i (DV^i) &=& 0 ~,  \label{eqn:de} \\
 \partial_t {\cal E} + \partial_i \left(-\sqrt{-g}~T^i_0\right) &=&
      -\sqrt{-g}~T^\alpha_\beta~\Gamma^\beta_{0 \alpha} ~,
    \label{eqn:en} \\
 \partial_t {\cal S}_j + \partial_i \left(\sqrt{-g}~T^i_j\right) &=&
      \sqrt{-g}~T^\alpha_\beta~\Gamma^\beta_{j \alpha} ~,
    \label{eqn:mom} \\
 \partial_t \mathcal{B}^j + \partial_i (\mathcal{B}^j V^i - \mathcal{B}^i V^j) &=&
     0 ~, 
      \label{eqn:ind} 
\end{eqnarray}
where $D=W\rho$ is the generalized fluid density, $W=\sqrt{-g} u^0 = \sqrt{-g}\gamma/\alpha$, $\gamma = 1/\sqrt{1-v^2}$, $\alpha = 1/\sqrt{-g^{00}}$, $v^2\equiv v_i v^i$, $v^i$ is the flow velocity relative to the normal observer, $V^i=u^i/u^0$ is the transport velocity, ${\cal E} = -\sqrt{-g} T^0_0$ is the total energy density, and ${\cal S}_j = \sqrt{-g} T^0_j$ is the covariant momentum density. With indices, $\Gamma$ indicates the geometric connection coefficients of the metric. Along with the evolution equations (\ref{eqn:de})-(\ref{eqn:ind}), the time component of the Maxwell equation enforces the divergence-free constraint $\partial_j \mathcal{B}^j = 0$.

There are multiple representations of the magnetic field in our equations, which we should explain: $b^\alpha$ is the magnetic field measured by an observer comoving with the fluid, which can be defined in terms of the dual of the Faraday tensor $b^\alpha \equiv u_\beta {^*F^{\alpha\beta}}$, and $\mathcal{B}^j = \sqrt{-g} B^j$ is the boosted magnetic field 3-vector. The magnetic field $B^i = {^*F^{\alpha i}}$ is related to the comoving field by 
\begin{equation}
B^i = u^0 b^i - u^i b^0 ~.
\end{equation}
We also need the magnetic pressure, which is defined as $P_\mathrm{mag} = b^2/2 = b^\alpha b_\alpha/2$. Note that unlike previous versions of {\em Cosmos++}, here we have absorbed the factor of $\sqrt{4\pi}$ into the definition of the magnetic fields.

\subsection{HRSC}
\label{sec:HRSC}
The current work uses a high-resolution shock capturing (HRSC) scheme to solve the GRMHD equations (\ref{eqn:de})-(\ref{eqn:ind}).  This method is new to {\em Cosmos++}, although it shares many elements with our earlier non-oscillatory central difference (NOCD) method \citep{anninos03}.  We begin by noting that the MHD equations are all in the form of a conservation equation
\begin{equation}
\partial_t \mathbf{U}(\mathbf{P}) + \partial_i \mathbf{F}^i(\mathbf{P}) = \mathbf{S}(\mathbf{P}) \label{eqn:cons}
\end{equation}
where
\begin{equation}
\mathbf{U}(\mathbf{P}) = \left( \begin{array}{c} D \\ {\cal E} \\ {\cal S}_j \\ \mathcal{B}^j \end{array} \right),~~ \mathbf{F}^i(\mathbf{P}) = \left( \begin{array}{c} DV^i \\  -\sqrt{-g}~T^i_0 \\ \sqrt{-g}~T^i_j \\  \mathcal{B}^j V^i - \mathcal{B}^i V^j \end{array} \right),~~ \mathbf{S}(\mathbf{P}) = \left( \begin{array}{c} 0 \\ -\sqrt{-g}~T^\alpha_\beta~\Gamma^\beta_{0 \alpha} \\ \sqrt{-g}~T^\alpha_\beta~\Gamma^\beta_{j \alpha} \\ 0 \end{array} \right)
\end{equation}
are the arrays representing the conserved quantities, fluxes, and source terms, respectively. By integrating both sides of equation (\ref{eqn:cons}) with respect to volume and applying Gauss' Theorem, we can rewrite the conservation equation in the form
\begin{equation}
\int_V \partial_t \mathbf{U} dV = -\oint_S \mathbf{F}^i dA_i + \int_V \mathbf{S} dV ~,
\end{equation} 
where we have dropped the explicit dependence on the primitive variables. We can then discretize this equation using a finite volume representation as
\begin{equation}
\mathbf{U}^{n+1} = \mathbf{U}^n - \frac{\Delta t}{V} \sum\limits_{faces}\left(\mathbf{F}^i A_i\right) + \Delta t \mathbf{S} ~.
\label{eqn:consstep}
\end{equation}
This approach requires at least a 2nd order time integration scheme for stability.  The {\em Cosmos++} code has a number of  time integration options, including: 2nd order Euler, 2nd order Runge-Kutta, 2nd order Crank-Nicholson, and 3rd order Euler.  The present work uses the 2nd order Runge-Kutta scheme (which is also sometimes referred to as the ``midpoint'' or ``leapfrog'' method). First a half-time step, $\Delta t/2$, is taken to project the conserved variables $\mathbf{U}^n$ forward to $n+1/2$. From these, a new set of primitives $\mathbf{P}^{n+1/2}$ can be computed. These intermediate primitives are then used in calculating the $\mathbf{F}^i$ and $\mathbf{S}$ needed in equation (\ref{eqn:consstep}).

The $\mathbf{F}^i$ are determined using an approximate Riemann solver. We have options for either the HLL or Lax-Friedrich method. The HLL scheme reconstructs the fluxes as  
\begin{equation}
\mathbf{F} = \frac{c_{min} \mathbf{F}_R + c_{max} \mathbf{F}_L - c_{max} c_{min} ( \mathbf{U}_R - \mathbf{U}_L )}{c_{max} + c_{min}} ~.
\end{equation}
A slope-limited linear or parabolic extrapolation gives $\mathbf{P}_R$ and $\mathbf{P}_L$, the primitive variables at the right- and left-hand side of each zone interface.  From $\mathbf{P}_R$ and $\mathbf{P}_L$, we calculate the right- and left-hand conserved quantities ($\mathbf{U}_R$ and $\mathbf{U}_L$), the fluxes $\mathbf{F}_R = \mathbf{F}(\mathbf{P}_R)$ and $\mathbf{F}_L = \mathbf{F}(\mathbf{P}_L)$, and the maximum right- and left-going waves speeds, $c_{\pm , R}$ and $c_{\pm , L}$. The bounding wave speeds are then $c_\mathrm{max} \equiv \mathrm{max}(0, ~c_{+,R}, ~c_{+,L})$ and $c_\mathrm{min} \equiv -\mathrm{min}(0, ~c_{-,R}, ~c_{-,L})$. By setting $c_{max} = c_{min}$, the HLL flux can be reduced to the local Lax-Friedrichs flux.

For the parabolic interpolations of $\mathbf{P}_R$ and $\mathbf{P}_L$, we use the piecewise-parabolic method (PPM) of \citet{colella84}.  As noted by those authors, this method can, on occasion, produce unwanted oscillations, especially behind stationary shocks, such as the ones to be treated in \S \ref{sec:tests}.  To combat these oscillations, we added the flattening procedure described in \citet{colella84}.

As noted above, the conserved variables $\mathbf{U}$, the fluxes $\mathbf{F}^i$, and the source terms $\mathbf{S}$ are all functions of the set of primitive variables
\begin{equation}
\mathbf{P} = \left( \begin{array}{c} \rho \\ \rho\epsilon \\ V^j \\ B^j \end{array} \right) ~.
\end{equation}
One difficulty in relativistic MHD is the so-called ``inversion problem,'' that is going from updated conserved variables to updated primitive variables. Unlike Newtonian MHD, there is not a set of analytically-solvable algebraic expressions for this inversion. Instead, one must use a numerical procedure, such as the Newton-Raphson method, to solve one or more of the inversion equations.  In our code we have implemented the $2D$, $1D_W$, and $1D_{v^2}$ methods of \citet{noble06}.  Our code defaults to the $2D$ method, but will fall back to the other methods in succession if the $2D$ method fails to converge.

\subsection{Constrained Transport}
\label{sec:CT}
The finite-volume discretization of the induction equation presented in \S \ref{sec:HRSC} can be treated using the constrained transport schemes of \citet{toth00}. However, these methods have certain inadequacies we wish to avoid. Most importantly for us, they are not easily extendible to adaptive mesh refinement. Another significant shortcoming is the rather large stencil they require. The methods can also lead to unphysical behavior for certain types of flows \citep{gardiner08}.

Instead, we depart from the volume-averaged representation of all of the fields to use a staggered representation of the magnetic fields, with the primary representation of the fields being face-centered. We then integrate the induction equation (\ref{eqn:ind}) over surfaces rather than the volume of the cell. For example, integrating equation (\ref{eqn:ind}) over the $x_1$-, $x_2$-, and $x_3$-faces located at $(i-1/2,j,k)$, $(i,j-1/2,k)$, and $(i,j,k-1/2)$ respectively gives, after use of StokeÕs Law, \citep{stone09}
\begin{eqnarray}
(\mathcal{B}^1)^{n+1}_{i-1/2,j,k} & = & (\mathcal{B}^1)^n_{i-1/2,j,k} - \frac{\delta t}{\delta x_2} \left[ \left(\mathscr{E}_3\right)^{n+1/2}_{i-1/2,j+1/2,k} - \left(\mathscr{E}_3\right)^{n+1/2}_{i-1/2,j-1/2,k} \right] \\ \nonumber
 & & +\frac{\delta t}{\delta x_3} \left[ \left(\mathscr{E}_2\right)^{n+1/2}_{i-1/2,j,k+1/2} - \left(\mathscr{E}_2\right)^{n+1/2}_{i-1/2,j,k-1/2} \right] ~,\\
(\mathcal{B}^2)^{n+1}_{i,j-1/2,k} & = & (\mathcal{B}^2)^n_{i,j-1/2,k} - \frac{\delta t}{\delta x_3} \left[ \left(\mathscr{E}_1\right)^{n+1/2}_{i,j-1/2,k+1/2} - \left(\mathscr{E}_1\right)^{n+1/2}_{i,j-1/2,k-1/2} \right] \\ \nonumber
 & & +\frac{\delta t}{\delta x_1} \left[ \left(\mathscr{E}_3\right)^{n+1/2}_{i+1/2,j-1/2,k} - \left(\mathscr{E}_3\right)^{n+1/2}_{i-1/2,j-1/2,k} \right] ~,\\
(\mathcal{B}^3)^{n+1}_{i,j,k-1/2} & = & (\mathcal{B}^3)^n_{i,j,k-1/2} - \frac{\delta t}{\delta x_1} \left[ \left(\mathscr{E}_2\right)^{n+1/2}_{i+1/2,j,k-1/2} - \left(\mathscr{E}_2\right)^{n+1/2}_{i-1/2,j,k-1/2} \right] \\ \nonumber
 & & +\frac{\delta t}{\delta x_2} \left[ \left(\mathscr{E}_1\right)^{n+1/2}_{i,j+1/2,k-1/2} - \left(\mathscr{E}_1\right)^{n+1/2}_{i,j-1/2,k-1/2} \right] ~,
\end{eqnarray}
where, for example, $(\mathscr{E}_1)_{i,j-1/2,k-1/2}$ is the $x_1$-component of the electric field (or emf) $\mathscr{E}_i = -\epsilon_{ijk} V^j \mathcal{B}^k$ centered on the appropriate cell edge. The best method we have found so far for constructing the edge-centered electric fields is to simply average the surrounding face-centered fluxes recovered from the Riemann solver. For example,
\begin{equation}
\left(\mathscr{E}_3\right)_{i-1/2,j-1/2,k} = \frac{1}{4} \left[ \left(\mathscr{E}_3\right)_{i-1/2,j,k} + \left(\mathscr{E}_3\right)_{i-1/2,j-1,k} + \left(\mathscr{E}_3\right)_{i,j-1/2,k} + \left(\mathscr{E}_3\right)_{i-1,j-1/2,k} \right] ~.
\end{equation}
This procedure preserves the following mathematical representation of the divergence to round-off error in each zone:
\begin{eqnarray}
\partial_i \mathcal{B}^i & = & \frac{1}{V} \left[ (\mathcal{B}^1 A_1)_{i+1/2,j,k} - (\mathcal{B}^1 A_1)_{i-1/2,j,k} + (\mathcal{B}^2 A_2)_{i,j+1/2,k} \right. \nonumber \\
 & & \left. - (\mathcal{B}^2 A_2)_{i,j-1/2,k} + (\mathcal{B}^3 A_3)_{i,j,k+1/2} - (\mathcal{B}^3 A_3)_{i,j,k-1/2} \right] ~.
\end{eqnarray}

In this scheme we maintain staggered forms of both the primitive and conserved magnetic fields, which are related by the metric determinant at their respective faces, e.g. $\mathcal{B}^1 = (\sqrt{-g})_{i-1/2,j,k} B^1$. The staggered primitive field is used to overwrite the appropriate component of the extrapolated primitive field used in the flux reconstruction at each face. They are also used when we require cell-centered values of the primitive magnetic fields, such as for calculating the magnetic pressure. In this case we use the volume-averaged fields
\begin{eqnarray}
(B^1)_{i,j,k} & = & \frac{1}{2} \left[ (B^1)_{i-1/2,j,k} + (B^1)_{i+1/2,j,k} \right] \\
(B^2)_{i,j,k} & = & \frac{1}{2} \left[ (B^2)_{i,j-1/2,k} + (B^2)_{i,j+1/2,k} \right] \\
(B^3)_{i,j,k} & = & \frac{1}{2} \left[ (B^3)_{i,j,k-1/2} + (B^3)_{i,j,k+1/2} \right] ~.
\end{eqnarray}

\subsection{Radiation}
We now describe the addition of the radiation field.  Similar to the fluid and magnetic field, the energy and momentum of the radiation field are represented by a stress-energy tensor
\begin{equation}
R^{\alpha \beta} = \int I_\nu n^\alpha n^\beta d\nu d\Omega ~,
\end{equation}
where $\nu$ is the photon frequency, $I_\nu = I(x^\alpha; n^i,\nu)$ is the specific intensity of radiation at position $x^\alpha$ moving in the direction $n^\alpha \equiv p^\alpha/h_P \nu$, $p^\alpha$ is the photon 4-momentum, $h_P$ is Planck's constant, and $d\Omega$ is the differential solid angle around the direction of propagation.  The quantities $\nu$, $I_\nu$, and $d\Omega$ are measured in the local Lorentz frame of an observer with 4-velocity $u^\alpha_\mathrm{(fid)}$.

GR {\em radiation} MHD, then, starts from the same conservation of stress-energy tensor as GRMHD, but now split into MHD and radiation components:
\begin{equation}
\left( T^{\alpha \beta} + R^{\alpha \beta} \right)_{;\beta} = 0 ~.
\label{eq:cons_stress}
\end{equation}
It is more convenient for our purposes to rewrite equation (\ref{eq:cons_stress}) in two parts:
\begin{equation}
\left( T^{\alpha \beta}\right)_{;\beta} = G^\alpha
\label{eqn:Tmhd}
\end{equation}
and
\begin{equation}
\left( R^{\alpha \beta}\right)_{;\beta} = -G^\alpha ~,
\label{eqn:Trad}
\end{equation}
where $G^\alpha$ is the radiation 4-force density coupling the fluid and radiation field.

\subsubsection{Radiation Moments}
Similar to fluid dynamics, we can proceed by considering various moments of the radiation equations.  Starting with the zeroth radiation moment, 
\begin{equation}
E = \int I_\nu d\nu d\Omega
\end{equation}
is the comoving radiation energy density, 
\begin{equation}
F^{\hat{i}} = \int I_\nu n^{\hat{i}} d\nu d\Omega
\end{equation}
is the comoving radiation flux (three first radiation moments), and
\begin{equation}
{\cal P}^{\hat{i} \hat{j}} = \int I_\nu n^{\hat{i}} n^{\hat{j}} d\nu d\Omega
\end{equation}
is the (symmetric) comoving radiation pressure tensor (six second radiation moments).  In the comoving frame, the radiation stress tensor is then
\begin{equation}
R^{\hat{\alpha}\hat{\beta}} = \left( \begin{array}{cccc} E & F^{\hat{1}} & F^{\hat{2}} & F^{\hat{3}} \\ F^{\hat{1}} & {\cal P}^{\hat{1}\hat{1}} & {\cal P}^{\hat{1}\hat{2}} & {\cal P}^{\hat{1}\hat{3}} \\  F^{\hat{2}} & {\cal P}^{\hat{2}\hat{1}} & {\cal P}^{\hat{2}\hat{2}} & {\cal P}^{\hat{2}\hat{3}} \\  F^{\hat{3}} & {\cal P}^{\hat{3}\hat{1}} & {\cal P}^{\hat{3}\hat{2}} & {\cal P}^{\hat{3}\hat{3}} \end{array} \right) ~.
\end{equation}
In our current implementation, we assume the radiation pressure, $P_\mathrm{rad}$, is isotropic, such that ${\cal P}^{\hat{i} \hat{j}} = \delta^{\hat{i} \hat{j}} P_\mathrm{rad} = \delta^{\hat{i} \hat{j}} E/3$.  This is equivalent to keeping the first two radiative moment equations, and adopting an Eddington factor of $1/3$ to close the set.  Although the radiation pressure is assumed to be isotropic, we emphasize that by allowing a small, nonzero radiation flux ($F^\alpha \ll E$), we do allow for some degree of anisotropy in the radiation field.  In covariant form, $R^{\alpha \beta}$ can now be written as
\begin{equation}
R^{\alpha \beta} = E u^\alpha u^\beta + F^\alpha u^\beta + F^\beta u^\alpha + P_\mathrm{rad} h^{\alpha \beta} ~,
\end{equation}
where we have introduced the tensor $h^{\alpha \beta} = g^{\alpha \beta} + u^\alpha u^\beta$, which projects any other tensor into the space orthogonal to $u^\alpha$, such that $h^{\alpha \beta} u_\alpha = 0$, and the radiation flux 4-vector
\begin{equation}
F^\alpha = \left. h^\alpha\right._\beta \int I_\nu n^\beta d\nu d\Omega ~.
\end{equation}
With this definition, the flux satisfies $F^\alpha u_\alpha = 0$.

\subsubsection{Radiation Four-Force Density}
\label{sec:four-force}

We return now to consider the form of the radiation 4-force density from equations (\ref{eqn:Tmhd}) and (\ref{eqn:Trad}).  In the comoving frame
\begin{equation}
G^{\hat{\alpha}} = \int \left(\chi_\nu I_\nu - \eta_\nu \right) n^{\hat{\alpha}} d\nu d\Omega ~,
\end{equation}
where $\chi_\nu = \chi_\nu^\mathrm{a} + \chi_\nu^\mathrm{s}$ is the total opacity (``$\mathrm{a}$'' and ``$\mathrm{s}$'' stand for absorption and scattering, respectively) and $\eta_\nu = \eta_\nu^\mathrm{a} + \eta_\nu^\mathrm{s}$ is the total emissivity.  In terms of the usual cooling function $\Lambda = \int \eta_\nu d\nu d\Omega$ and assuming a mean (frequency-independent) opacity of the form $\chi = \rho\kappa$, we can write in the comoving frame
\begin{equation}
G^{\hat{0}} = \int \left(\chi_\nu I_\nu - \eta_\nu \right) d\nu d\Omega = \rho \kappa^\mathrm{a} E - \Lambda ~,
\label{eq:G0}
\end{equation}
and 
\begin{equation}
G^{\hat{i}} = \int \left(\chi_\nu I_\nu - \eta_\nu \right) n^{\hat{i}} d\nu d\Omega = \rho \kappa F^{\hat{i}}  = \rho (\kappa^\mathrm{a} + \kappa^\mathrm{s}) F^{\hat{i}} ~.
\label{eq:Gi}
\end{equation}
Equation (\ref{eq:Gi}) assumes that photons are emitted isotropically in the comoving frame, so that the net momentum they remove from the gas is zero.  

Noting that $u^{\hat{\alpha}} = (1, 0, 0, 0)$, then, from the normalization of the radiation flux 4-vector $F^\alpha u_\alpha = 0$, we have $F^{\hat{\alpha}} = (0, F^{\hat{i}})$.  Thus, in the comoving frame
\begin{equation}
G^{\hat{\alpha}} = \left(\rho \kappa^\mathrm{a} E - \Lambda \right) u^{\hat{\alpha}} + \rho \kappa F^{\hat{\alpha}} ~.
\label{eq:Galpha}
\end{equation}
Since this expression is covariant, it must hold in any frame; we, therefore, drop the $\hat{ }$ in all subsequent references to $G^\alpha$.  

For equilibrium blackbody radiation, $\Lambda = \rho \kappa^\mathrm{a} a_R T_\mathrm{gas}^4$, where $T_\mathrm{gas} = m P_\mathrm{gas}/k_B\rho$ is the ideal gas temperature of the fluid, $a_R=8\pi^5 k_B^4/(15 h_p^3 c^3)$ is the radiation constant, $k_B$ is Boltzmann's constant, and $m = \mu m_\mathrm{H}$ is the mean mass of ions in the gas.  If the radiation and gas are in local thermodynamic equilibrium, then $E = a_R T_\mathrm{gas}^4$ and the first two terms of equation (\ref{eq:Galpha}) drop out, though we emphasize that our scheme does not require this.

\subsubsection{GR Radiation MHD}
Again expanding and reorganizing the conservation of stress-energy in terms of new variables, we get four new conservation equations for the radiation
\begin{eqnarray}
 \partial_t {\cal R} + \partial_i \left(\sqrt{-g}~R^i_0\right) &=&
      \sqrt{-g}~R^\alpha_\beta~\Gamma^\beta_{0 \alpha} - \sqrt{-g}~G_0
    \label{eqn:rad_en} \\
 \partial_t {\cal R}_j + \partial_i \left(\sqrt{-g}~R^i_j\right) &=&
      \sqrt{-g}~R^\alpha_\beta~\Gamma^\beta_{j \alpha} - \sqrt{-g}~G_j ~,
    \label{eqn:rad_mom} 
\end{eqnarray}
where ${\cal R} = \sqrt{-g}R^0_0$ is the conserved radiation energy density and ${\cal R}_j = \sqrt{-g} R^0_j$ is the conserved radiation momentum density.  The addition of the radiation fields also modifies the MHD energy-momentum conservation equations.  The new equations are
\begin{eqnarray}
 \partial_t {\cal E} + \partial_i \left(-\sqrt{-g}~T^i_0\right) &=&
      -\sqrt{-g}~T^\alpha_\beta~\Gamma^\beta_{0 \alpha} - \sqrt{-g}~G_0 ~,
    \label{eqn:en2} \\
 \partial_t {\cal S}_j + \partial_i \left(\sqrt{-g}~T^i_j\right) &=&
      \sqrt{-g}~T^\alpha_\beta~\Gamma^\beta_{j \alpha} + \sqrt{-g}~G_j ~,
    \label{eqn:mom2}
\end{eqnarray}

The full set of conserved variables, fluxes, and source terms is now
\begin{equation}
\mathbf{U}(\mathbf{P}) = \left( \begin{array}{c} D \\ {\cal E} \\ {\cal S}_j \\ \mathcal{B}^j \\ {\cal R} \\ {\cal R}_j \end{array} \right),~~ \mathbf{F}^i(\mathbf{P}) = \left( \begin{array}{c} DV^i \\  -\sqrt{-g}~T^i_0 \\ \sqrt{-g}~T^i_j \\  \mathcal{B}^j V^i - \mathcal{B}^i V^j \\ \sqrt{-g}~R^i_0 \\ \sqrt{-g}~R^i_j \end{array} \right),~~ \mathbf{S}(\mathbf{P}) = \left( \begin{array}{c} 0 \\ -\sqrt{-g}~T^\alpha_\beta~\Gamma^\beta_{0 \alpha} - \sqrt{-g}~G_0 \\ \sqrt{-g}~T^\alpha_\beta~\Gamma^\beta_{j \alpha} + \sqrt{-g}~G_j \\ 0 \\ \sqrt{-g}~R^\alpha_\beta~\Gamma^\beta_{0 \alpha} - \sqrt{-g}~G_0 \\ \sqrt{-g}~R^\alpha_\beta~\Gamma^\beta_{j \alpha} - \sqrt{-g}~G_j \end{array} \right) ~.
\end{equation}
The expanded set of primitive variables is now
\begin{equation}
\mathbf{P} = \left( \begin{array}{c} \rho \\ \rho\epsilon \\ V^j \\ B^j \\ E \\ F^j \end{array} \right) ~.
\end{equation}
Fortunately the inversion of the radiation variables can be written as a simple set of algebraic expressions (utilizing $P_\mathrm{rad} = E/3$), and can be solved independently of the MHD primitive fields:
\begin{eqnarray}
E & = & 3 \left( \frac{{\cal R}^0 + 2 u^0 u_\alpha {\cal R}^\alpha}{\sqrt{-g} g^{00} - 2 W u^0} \right) \\
F^0 & = & \frac{-1}{\sqrt{-g}} \left( WE + u_\alpha {\cal R}^\alpha \right) \\
F^j & = & \frac{ {\cal R}^j}{W} - F^0 \frac{u^j}{u^0} - \frac{4}{3} E u^j - \frac{g^{0j} E}{3 u^0} ~.
\end{eqnarray}
Inside the code we check that the radiation flux satisfies the physical limit $\sqrt{F_\mu F^\mu} \le E$.

The advantage of this approach is that, by writing the radiation equations in conservative form, we can take full advantage of the HRSC machinery described in \S \ref{sec:HRSC} to solve the full set of GR radiation MHD equations.  Thus, there is relatively little code development beyond adding new variables, flux terms, and source terms.

\subsection{Wave Speed Calculation}
The HLL and Lax-Friedrich approximate Riemann solvers only require the maximum and minimum wave speeds, as opposed to the full eigenvectors of the characteristic matrix as would be necessary for a Roe-type scheme. These wave speeds are also required to fix the time step via the Courant condition. The relevant speed is the phase speed $\omega/k$ of the wave, and we treat each coordinate direction independently. For signals propagating in the $x_1$ coordinate direction, the corresponding eigenvector is $k_\alpha = (-\omega, k_1, 0 , 0)$, and the wave speed is $\omega/k_1$.

To find the necessary wave speeds in the grid frame, we start with the following approximate dispersion relation in the comoving frame \citep{gammie03,farris08}
\begin{equation}
\omega^2_\mathrm{cm} = v^2_\mathrm{T} k^2_\mathrm{cm} ~,
\label{eqn:dispersion}
\end{equation}
where
\begin{equation}
v^2_\mathrm{T} = \mathrm{max} \left\{ \begin{array}{l} 1/3 \\ v_\mathrm{A}^2 + c_\mathrm{s}^2\left(1-v_\mathrm{A}^2\right) \end{array} \right. ~,
\end{equation}
$v_\mathrm{A}^2 = b^2/(\rho h + b^2)$ is the Alfv\'en speed, and $c_\mathrm{s}^2 = \Gamma P_\mathrm{gas}/\rho h$ is the sound speed. We then substitute the following relations, $\omega_\mathrm{cm} = -k_\alpha u^\alpha$, $k_\mathrm{cm}^2 = K_\alpha K^\alpha$, and $K_\alpha = (g_{\alpha\beta} + u_\alpha u_\beta)k^\beta$, into equation (\ref{eqn:dispersion}) to get the following quadratic equation for the desired wave speed $w/k_1$
\begin{eqnarray}
\left[1 - v_T^2\left(1 + \frac{g^{00}}{u^0u^0}\right)\right]\left(\frac{\omega}{k_1}\right)^2 + 2\left[v_T^2\left(V^1 + \frac{g^{01}}{u^0u^0}\right) -V^1\right]\left(\frac{\omega}{k_1}\right) + \\ \nonumber
\left[V^1V^1 - v_T^2\left(V^1V^1 + \frac{g^{11}}{u^0u^0}\right)\right] = 0 ~.
\end{eqnarray}
Wave speeds in the $x_2$ and $x_3$ directions are found analogously.

\section{Radiation shock tube tests}
\label{sec:tests}

Unfortunately, there are very few good test problems for relativistic radiation MHD codes at this time.  Among the few are the four radiative shock tube tests introduced in \citet{farris08}, which we now use to validate our code.  Each test includes a nonlinear radiation-hydrodynamic wave:  case 1 is a nonrelativistic strong shock; case 2 is a mildly relativistic strong shock; case 3 is a highly relativistic wave; and case 4 is a radiation-pressure-dominated, mildly relativistic wave.  

Similar to \citet{zanotti11}, and unlike \citet{farris08}, we initiate these problems as traditional shock tubes with two states (``Left'' and ``Right''), initially separated by an imaginary partition, instead of starting from the analytic solution.  At $t=0$ the partition is removed and the gas and radiation are allowed to evolve until a steady state is reached.  The initial parameters for the four tests are presented in Table \ref{tab:shocktube}.  The initial values for the fluxes, not given in Table \ref{tab:shocktube}, are set to $F^x = 10^{-2} E$.  The scattering opacity $\kappa^\mathrm{s}$ is set to zero in all these tests.  Also like \citet{zanotti11} and unlike \citet{farris08}, we do not boost the fluid velocities, instead presenting our results in the frame of the shock.  This is important, as quasi-stationary shocks present a particular difficulty for the PPM scheme we are using.  In fact, without using the flattening procedure of \citet{colella84}, we find that the two tests that exhibit discontinuities (cases 1 and 2), suffer from post-shock oscillations with amplitudes of a few percent.  With the flattening procedure, the oscillations are effectively removed, without creating unnecessary dissipation in the smooth test problems.

\begin{deluxetable}{cccccccccccc}
\tabletypesize{\scriptsize}
\tablecaption{Shock Tube Parameters \label{tab:shocktube}}
\tablewidth{0pt}
\tablehead{
\colhead{Case} & \colhead{$\Gamma$} & \colhead{$\kappa^\mathrm{a}$} &
\colhead{$\rho_L$} & \colhead{$P_L$} & \colhead{$u^x_L$} & \colhead{$E_L$} & \colhead{$\rho_R$} & \colhead{$P_R$} & \colhead{$u^x_R$} & \colhead{$E_R$} & \colhead{$t_\mathrm{stop}$}
}
\startdata
1 & 5/3 & 0.4 & 1.0 & $3.0 \times 10^{-5}$ & 0.0015 & $1.0 \times 10^{-8}$ & 2.4 & $1.61 \times 10^{-4}$ & $6.25 \times 10^{-3}$ & $2.51 \times 10^{-7}$ & 4000 \\
2 & 5/3 & 0.2 & 1.0 & $4.0 \times 10^{-3}$ & 0.25 & $2.0 \times 10^{-5}$ & 3.11 & 0.04512 & 0.0804 & $3.46 \times 10^{-3}$ & 3000 \\
3 & 2 & 0.3 & 1.0 & 60.0 & 10.0 & 2.0 & 8.0 & $2.34 \times 10^3$ & 1.25 & $1.14 \times 10^3$ & 100 \\
4 & 5/3 & 0.08 & 1.0 & $6.0 \times 10^{-3}$ & 0.69 & 0.18 & 3.65 & $3.59 \times 10^{-2}$ & 0.189 & 1.30 & 500
\enddata
\end{deluxetable}

\begin{figure}
\plotone{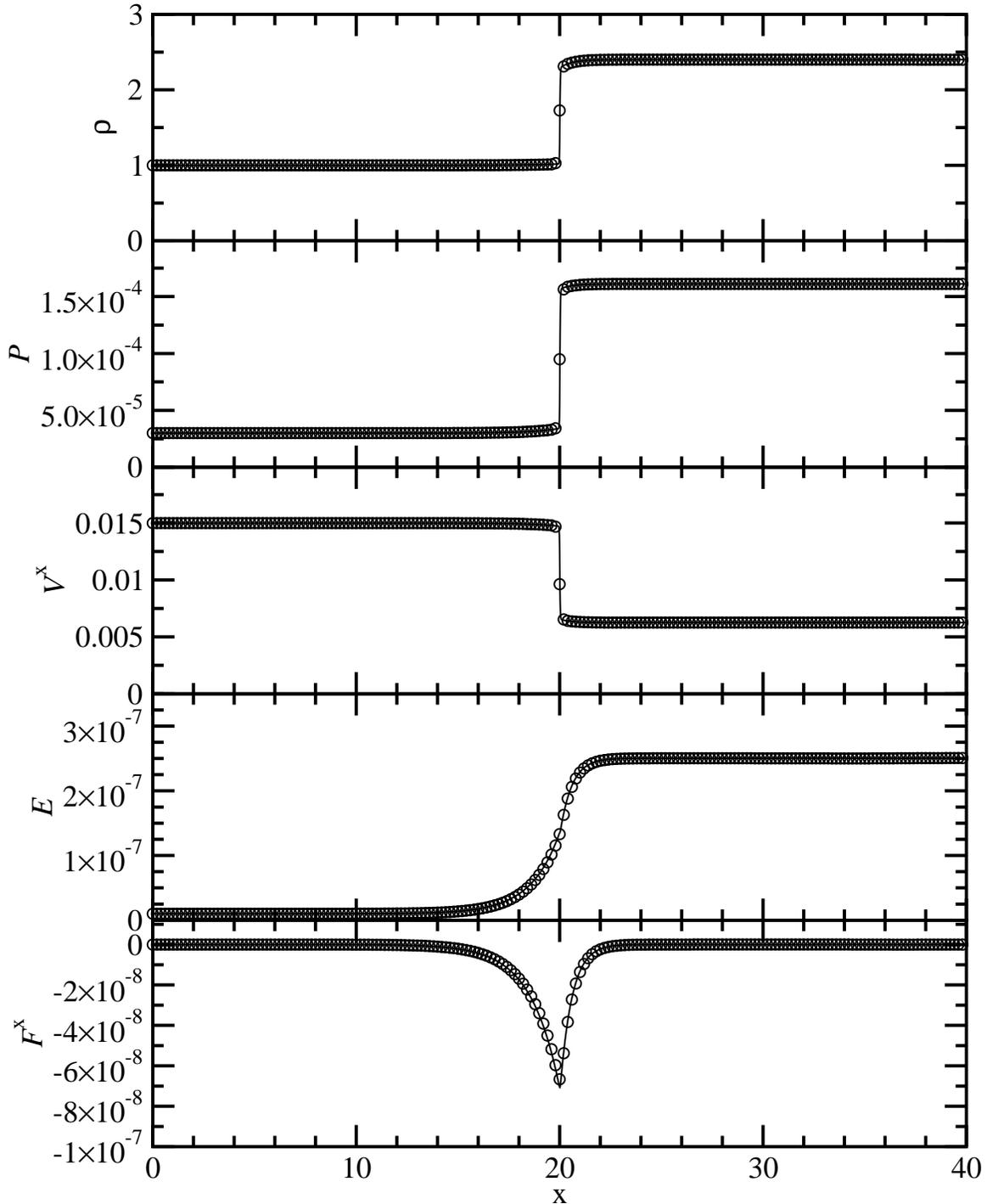}
\caption{Profiles of $\rho$, $P$, $V^x$, $E$, and $F^x$ at $t=4000$ for case 1.  Symbols denote data from 800-zone simulation (sampled to only display 200 points); solid lines denote data from 3200-zone simulation. 
\label{fig:case1}}
\end{figure}

\begin{figure}
\plotone{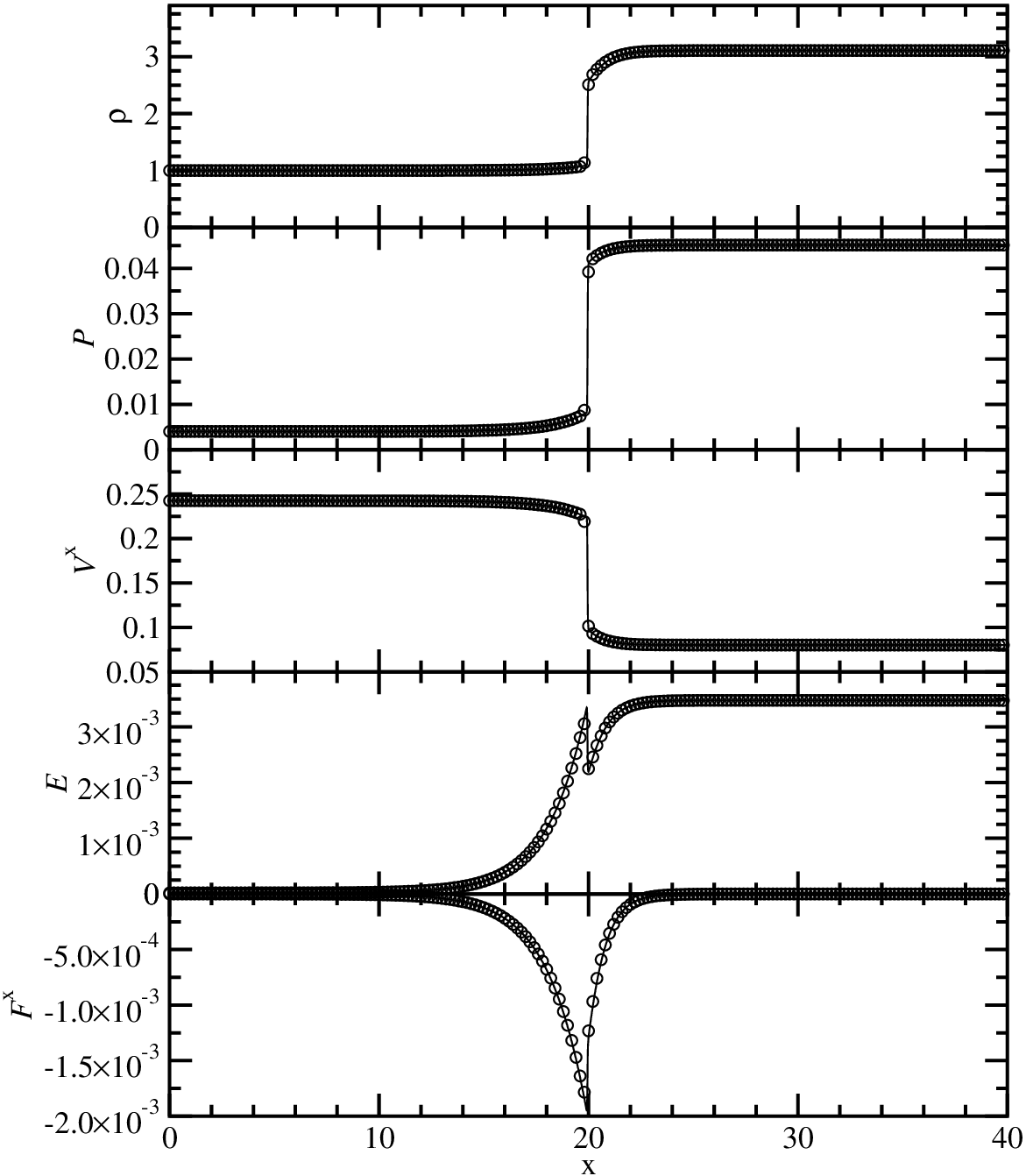}
\caption{Profiles of $\rho$, $P$, $V^x$, $E$, and $F^x$ at $t=3000$ for case 2.  Symbols denote data from 800-zone simulation (sampled to only display 200 points); solid lines denote data from 3200-zone simulation. 
\label{fig:case2}}
\end{figure}

\begin{figure}
\plotone{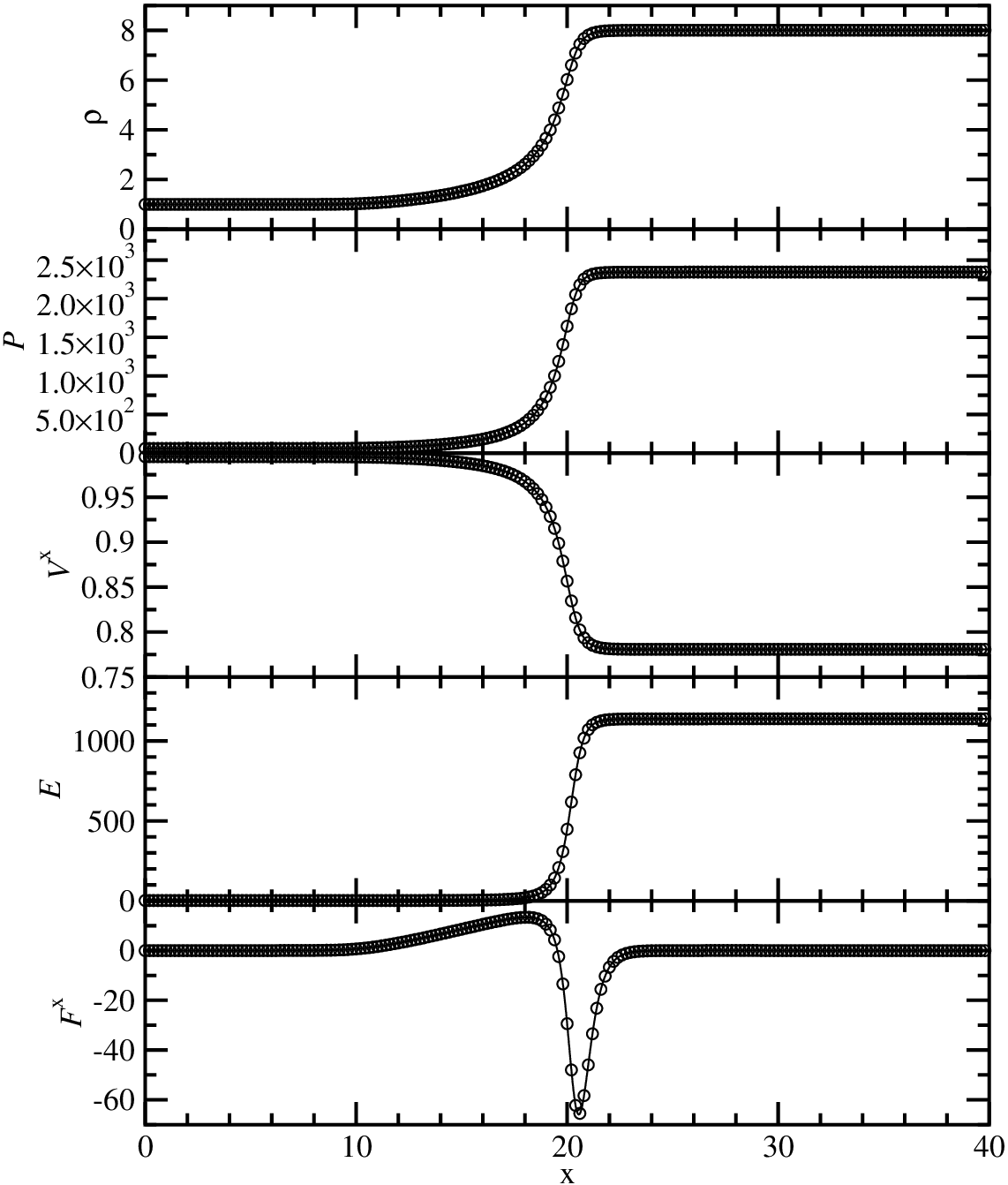}
\caption{Profiles of $\rho$, $P$, $V^x$, $E$, and $F^x$ at $t=100$ for case 3.  Symbols denote data from 800-zone simulation (sampled to only display 200 points); solid lines denote data from 3200-zone simulation. 
\label{fig:case3}}
\end{figure}

\begin{figure}
\plotone{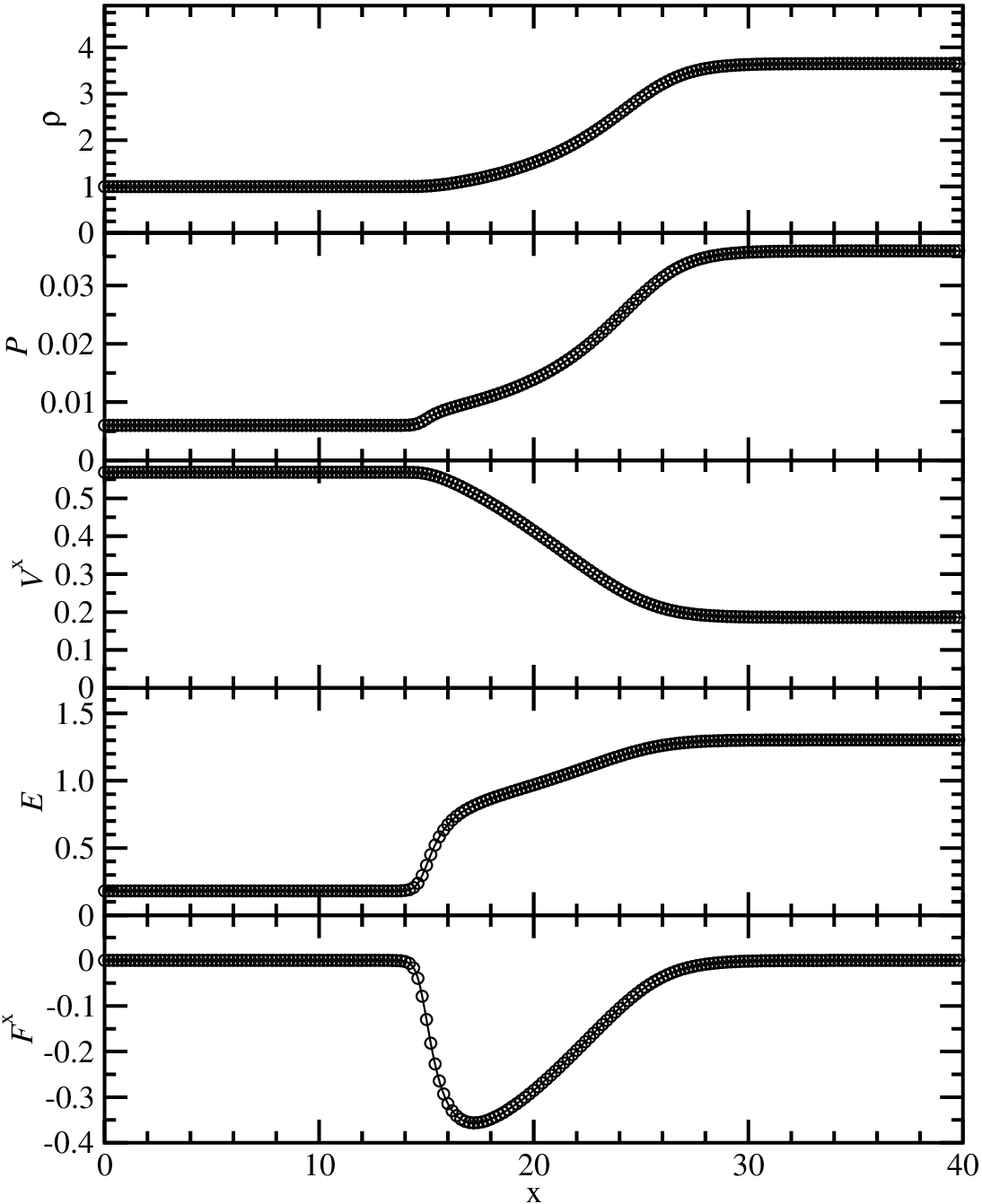}
\caption{Profiles of $\rho$, $P$, $V^x$, $E$, and $F^x$ at $t=500$ for case 4.  Symbols denote data from 800-zone simulation (sampled to only display 200 points); solid lines denote data from 3200-zone simulation. 
\label{fig:case4}}
\end{figure}

Figures \ref{fig:case1} -- \ref{fig:case4} show the four cases.  Although it is possible to calculate semi-analytic solutions for each of these \citep{farris08}, we have chosen instead to simply plot results using 800 zone resolution against results using a much higher (3200 zone) resolution.  In all cases, the results agree quite well at the different resolutions, and with previously published results.

We have used the smooth wave in case 4 to test the convergence rate of our numerical scheme.  Table \ref{tab:errors} reports the $L$-1 norm error (i.e. $\vert E(a) \vert_1 = \sum_i \Delta x \vert a_i - A_i \vert$, where $a_i$ and $A_i$ are the numerical and semi-analytic solutions, respectively) for 1600, 3200, and 6400 zones resolution.  The convergence rate of the errors for all variables at all resolutions is almost exactly 2, as expected for a smooth flow using our overall scheme.

\begin{deluxetable}{cccccc}
\tabletypesize{\scriptsize}
\tablecaption{$L$-1 Norm Errors for Case 4 \label{tab:errors}}
\tablewidth{0pt}
\tablehead{
\colhead{Grid} & \colhead{$\vert E(\rho) \vert_1$} & \colhead{$\vert E(P) \vert_1$} & \colhead{$\vert E(V^x) \vert_1$} & \colhead{$\vert E(E) \vert_1$} & \colhead{$\vert E(F^x) \vert_1$} 
}
\startdata
1600 & $1.69 \times 10^{-6}$ & $2.85 \times 10^{-8}$ & $1.17 \times 10^{-6}$ & $8.98 \times 10^{-7}$ & $1.17 \times 10^{-7}$  \\
3200 & $4.22 \times 10^{-7}$ & $7.06 \times 10^{-9}$ & $2.89 \times 10^{-7}$ & $2.23 \times 10^{-7}$ & $2.91 \times 10^{-8}$  \\
6400 & $1.06 \times 10^{-7}$ & $1.78 \times 10^{-9}$ & $7.16 \times 10^{-8}$ & $5.54 \times 10^{-8}$ & $7.27 \times 10^{-9}$ \\
\hline
Convergence\tablenotemark{a} & 1.99 & 1.99 & 2.01 & 2.01 & 2.00
\enddata
\tablenotetext{a}{Convergence rate between 3200 and 6400 zone data.}
\end{deluxetable}

\section{Bondi inflow with radiation}
\label{sec:bondi}

As we mentioned in \S \ref{sec:intro}, the case of optically-thick, spherical accretion onto a non-rotating black hole has been considered many times in the past, although usually with strictly one-dimensional codes.  Another difference is that most of the codes in the past have used implicit integration schemes, whereas we are, for now, attempting to proceed using the explicit scheme already in {\em Cosmos++}.  Finally, instead of assuming local thermodynamic equilibrium between the gas and the radiation, we consider two physical cooling processes in the gas: Thomson scattering and thermal bremsstrahlung.  The first contributes an opacity 
\begin{equation}
\kappa^\mathrm{s} = 0.4 ~\mathrm{cm}^2~\mathrm{g}^{-1} ~.
\end{equation}
while the second has the form  \citep{rybicki86} 
\begin{equation}
\kappa^\mathrm{a} = 1.7\times 10^{-25} T^{-7/2}_\mathrm{K} \rho_\mathrm{cgs} m^{-2}_\mathrm{p} ~\mathrm{cm}^2~\mathrm{g}^{-1} ~,
\end{equation}
where $T_\mathrm{K}$ is the ideal gas temperature of the fluid in Kelvin, $\rho_\mathrm{cgs}$ is the density in g/cm$^3$, and $m_\mathrm{p}$ is the mass of a proton in g.  We assume the gas is fully ionized hydrogen, so the mean molecular weight $\mu = 0.5$. 

We have chosen to set up cases similar to ones presented in \citet{vitello84} and \citet{nobili91}, though significant differences in our assumptions mean we do not expect to exactly reproduce their results.  In the case of \citet{vitello84}, they assume LTE throughout the flow, such that they do not need to solve the radiation transport independent of the fluid transport.  As we emphasized in \S \ref{sec:four-force}, we do not enforce LTE in our method.  There are even more differences between our approach and that of \citet{nobili91}.  First, they consider a more physical, though also more complicated, equation of state.  They also account for Compton scattering of the radiation, which we do not consider in this work.  Finally, they use a constant value of $T_o = 10^4$ K for the temperature of the gas at the outer boundary of their simulations, whereas we explore values in the range $10^5$--$10^7$ K.

We use a logarithmic radial coordinate of the form $x_1 \equiv 1+\ln(r/r_\mathrm{S})$ to cover the expansive spatial range required, where $r_\mathrm{S} = 2$ is the Schwarzschild radius.  All simulations use a resolution of 512 zones.  Note that because this problem is spherically symmetric, we have performed our tests using a 1D, spherical-polar (radial) grid.  Nevertheless, we have confirmed that our code produces identical results when the problem is run in 2 and 3 dimensions (when using the same radial grid spacing).

To initialize the problems, we first fix the density, $\rho_o$, and temperature, $T_o$, of the gas at $r_o$.  Once $T_o$ and $\rho_o$ are fixed, and assuming some relation between $T_\mathrm{gas}$ and $T_\mathrm{rad}$, we can determine the polytropic index of the gas at $r_o$ from
\begin{equation}
\Gamma = 1 + \frac{1}{3} \left(\frac{P_\mathrm{gas} + P_\mathrm{rad}}{P_\mathrm{gas}/2 + P_\mathrm{rad}} \right) ~.
\end{equation}
From this expression, we can easily see that $\Gamma=5/3$ for gas-pressure-dominated flows, whereas $\Gamma=4/3$ for radiation-pressure-dominated ones.  We assume the initial value of $\Gamma$ found at $r_o$ applies throughout the flow for the duration of the simulations.  By assuming a polytropic equation of state, $P \propto \rho^\Gamma$, we can determine the initial temperature profile of the gas from 
\begin{equation}
T_\mathrm{gas} = T_o (\rho/\rho_o)^{\Gamma-1} ~.
\label{eqn:T1}
\end{equation}
We still need to specify the initial profiles of $u^r$ and $\rho$.  For simplicity, we assume that these equal their free-fall values $u^r = -\sqrt{2M/r}$ and $\rho = -\dot{M}/4\pi r^2 u^r$ at all radii, with the mass accretion rate, $\dot{M}$, now one of our free parameters.  We explore mass accretion rates in the range $10 \le \dot{m} = \dot{M}/\dot{M}_\mathrm{Edd} \le 300$, where  $\dot{M}_\mathrm{Edd}$ is the Eddington mass accretion rate.  The lower limit is set by the requirement that at least some portion of the flow be optically thick.  The upper limit is set by the requirement that the ``photosphere'' be contained within the grid.  The photosphere is where the optical depth $\tau$ becomes 1, with $\tau$ defined as
\begin{equation}
\tau \equiv \int_0^X \left( \chi^\mathrm{a} + \chi^\mathrm{s} \right) ds ~.
\end{equation}
In practice we approximate the optical depth as $\tau \simeq (\chi^\mathrm{a} + \chi^\mathrm{s})r$.

We consider three types of simulations, all with $M = 3 M_\odot$, $r_o = 10^4 r_\mathrm{S}$, and $r_i = 0.95 r_\mathrm{S}$.  In the first, the gas is purely adiabatic, as we ignore radiation; in the second, we include radiative processes as described above; in the last, we also include a radial magnetic field of the form $\mathcal{B}^r = -
\partial_\theta A_\phi$, where
\begin{equation}
A_\phi = \mathrm{sign}(\cos \theta) \cos \theta \sqrt{2P_o/\beta_o} r_o^2 
\end{equation}
and $\beta_o = 100$.  Since the field is purely radial and weak, it has no noticeable effect on the dynamics.  Nevertheless, we wanted to include at least one test using all of the physics discussed in this paper.  Future work will rely much more heavily on the full capability.  

For the tests with radiation, we initially specify the ratio of radiation to gas temperature, $T_\mathrm{rad}/T_\mathrm{gas} \ll 1$.  This is mainly done so that the radiation energy density $E= a T^4_\mathrm{rad}$ may start with some reasonable value.  The radiation flux is similarly set to an arbitrary initial value $F^r \ll E$.  We confirm that our final results are not sensitive to our choices for these parameters.  The reason we initialize the radiation temperature to a much lower value than the gas is because, if the radiation and gas are nearly in thermal equilibrium, i.e. $E \approx a_R T_\mathrm{gas}^4$, then the first two terms of equation (\ref{eq:Galpha}) are approximately equal, meaning their difference can be arbitrarily small.  This makes the corresponding source terms of equations (\ref{eqn:rad_en})-(\ref{eqn:mom2}) incredibly stiff, and stable evolution with an explicit scheme would require an unacceptably small timestep.  In future work we plan to explore using an implicit scheme to solve this source term, which will alleviate this stability problem.  However, this limitation does not significantly affect the simulations presented here.  This is because the radiative processes we are considering are not very efficient, so the steady-state radiation temperature (and pressure) are naturally significantly less than the gas temperatures (and pressures) we consider.  

Table \ref{tab:thin} summarizes the key simulation parameters for this section.  Each simulation is run long enough for $E$ and $F^r$ to achieve steady-state profiles out to the photosphere.  Our main diagnostic in this section is the emitted luminosity, which can be recovered directly from the evolved radiation fluxes $F^i$.  Specifically,
\begin{equation}
L = \int_V F^i n_i dA ~,
\end{equation}
where $A$ is the surface area encompassing the volume $V$.  The surface is taken to correspond to the photosphere ($\tau = 1$).  We measure the luminosity in units of the Eddington luminosity $l = L/L_\mathrm{Edd}$, where $L_\mathrm{Edd} = \dot{M}_\mathrm{Edd} c^2 = 4 \pi G M c \sigma_T/m_p$ represents the limit at which outward radiation pressure balances gravity.  Above this limit, the radiation pressure is sufficient to halt, or even reverse, accretion.  
We define the radiative efficiency of our flows as $\eta = l/\dot{m}$.  

\begin{deluxetable}{cccccccc}
\tabletypesize{\scriptsize}
\tablecaption{Radiative Bondi Simulations \label{tab:thin}}
\tablewidth{0pt}
\tablehead{
\colhead{Simulation} & \colhead{$\dot{m}$} & \colhead{$T_o$ (K)} & \colhead{$P_\mathrm{rad}/P_\mathrm{gas}$\tablenotemark{a}} & \colhead{$t_\mathrm{stop}$} & cycles & $l$
}
\startdata
E10T5 & 10 & $10^5$ & $1.2\times10^{-7}$ & $7\times10^3$ & $7.8\times10^6$ & $2.01\times10^{-6}$ \\
E10a\tablenotemark{b} & 10 & $10^6$ & $\cdots$ & $7\times10^3$ & $2.7\times10^5$ & $\cdots$ \\
E10T6 & 10 & $10^6$ & $1.2\times10^{-4}$ & $7\times10^3$ & $7.8\times10^6$ & $3.72\times10^{-6}$ \\
E10T6b\tablenotemark{c} & 10 & $10^6$ & $1.2\times10^{-4}$ & $7\times10^3$ & $7.8\times10^6$ & $3.72\times10^{-6}$ \\
E10T7 & 10 & $10^7$ & 0.12 & $7\times10^3$ & $7.7\times10^6$ & $1.13\times10^{-5}$ \\
E30T6 & 30 & $10^6$ & $3.9\times10^{-5}$ & $7\times10^3$ & $7.7\times10^6$ & $2.66\times10^{-5}$ \\
E100T6 & 100 & $10^6$ & $1.2\times10^{-5}$ & $1.6\times10^4$ & $1.8\times10^7$ & $1.97\times10^{-4}$ \\
E300T6 & 300 & $10^6$ & $3.9\times10^{-6}$ & $4.3\times10^4$ & $4.8\times10^7$ & $7.91\times 10^{-4}$
\enddata
\tablenotetext{a}{Measured at the inner radial boundary at $t=0$.}
\tablenotetext{b}{Adiabatic}
\tablenotetext{c}{Radial magnetic field case}
\end{deluxetable}

\begin{figure}
\plotone{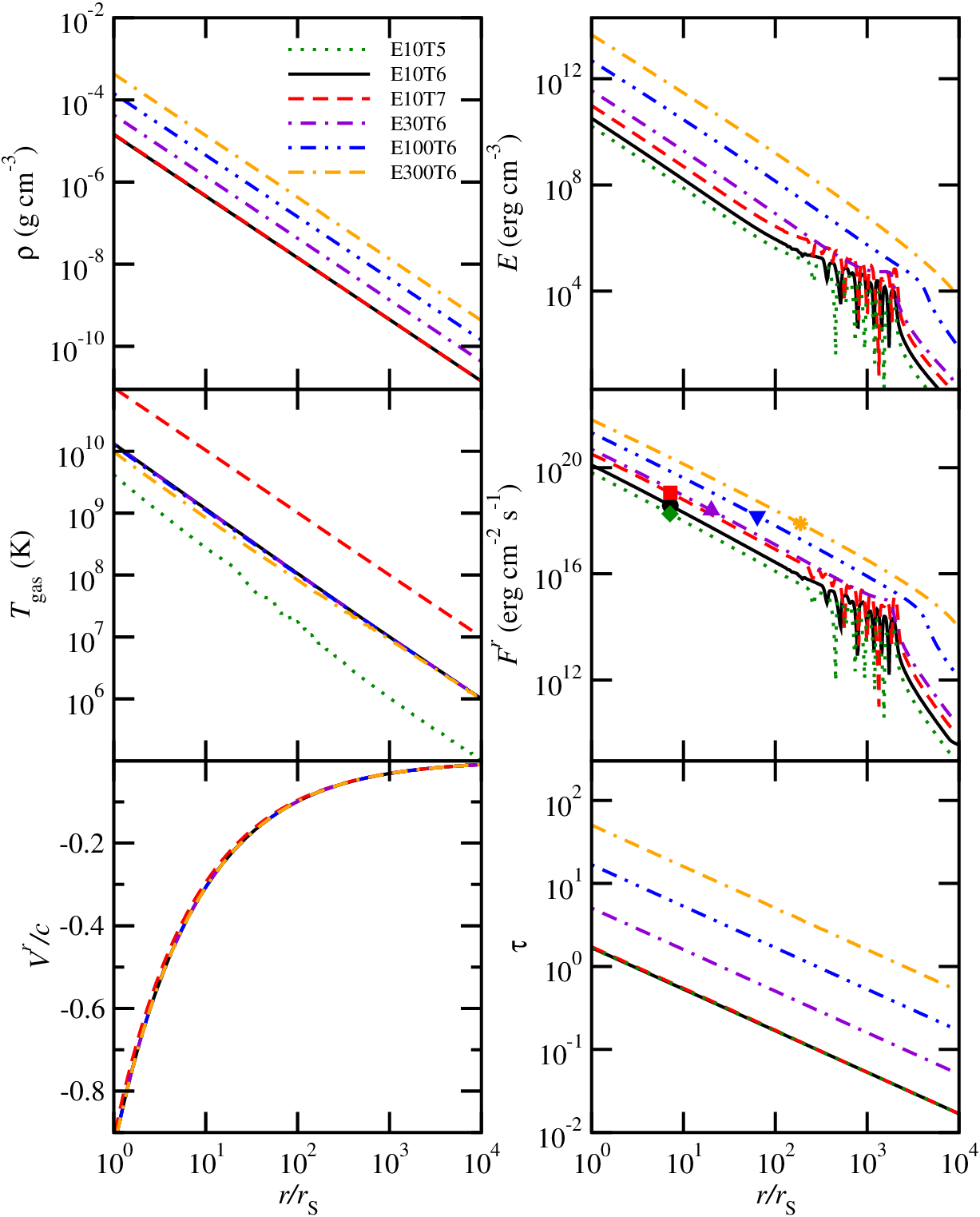}
\caption{Profiles of $\rho$, $T_\mathrm{gas}$, $V^r$, $E$, $F^r$, and $\tau$ for simulations with different values of $T_o$ and $\dot{m}$.  The symbols in the plot of $F^r$ denote the trapping radius for each case.  The lowest accretion rate cases, $\dot{m} \lesssim 10$, illustrates the stability problems that begin to plague our method when the flow becomes very optically thin.
\label{fig:profile}}
\end{figure}

Figure \ref{fig:profile} shows profiles of six different simulations exploring different values of $T_o$ and $\dot{m}$.  We did not include the adiabatic (E10a) nor radiation + magnetic field (E10T6b) cases in this plot since they are practically indistinguishable from our reference radiation simulation, E10T6.  The fact that the adiabatic and radiation simulations are so similar is not surprising since the radiation pressure is so much smaller than the gas pressure, $P_\mathrm{rad} < 0.025 P_\mathrm{gas}$ throughout the flow; so even with radiation, the flow behaves nearly adiabatically.  This reflects the very inefficient nature of the radiative processes we are considering.  The radiation + magnetic field simulations appears nearly identical, which again is expected since the magnetic field is purely radial and weak, so it can not play a dynamical role.  

In all of the cases the inflow is supersonic over the entire radial domain.  The accretion radius $r_a = GM/c^2_{s,\infty}$ also lies beyond the outer radial boundary for each of these cases.  We are, however, able to identify the trapping radius $r_t$, where the advection of photons inward becomes faster than their diffusion outward, i.e. $\vert E V^r \vert > F^r$, in all of our simulations; this point is marked with a symbol on each of the plots of $F^r$.

Simulations E10T5, E10T6, and E10T7 illustrate one of the main drawbacks of our current method - the restriction to optically thick flows.  For simulations with $\dot{m} \lesssim 10$, the flow is only optically thick in the inner few $r_\mathrm{S}$.  Once the optical depth drops below $\tau \approx 0.1$, we begin to see oscillations or noise in the profiles of $E$ and $F^r$.  These oscillations do not appear to damp away with time, and indicate a fundamental limit to our method.

\begin{figure}
\plotone{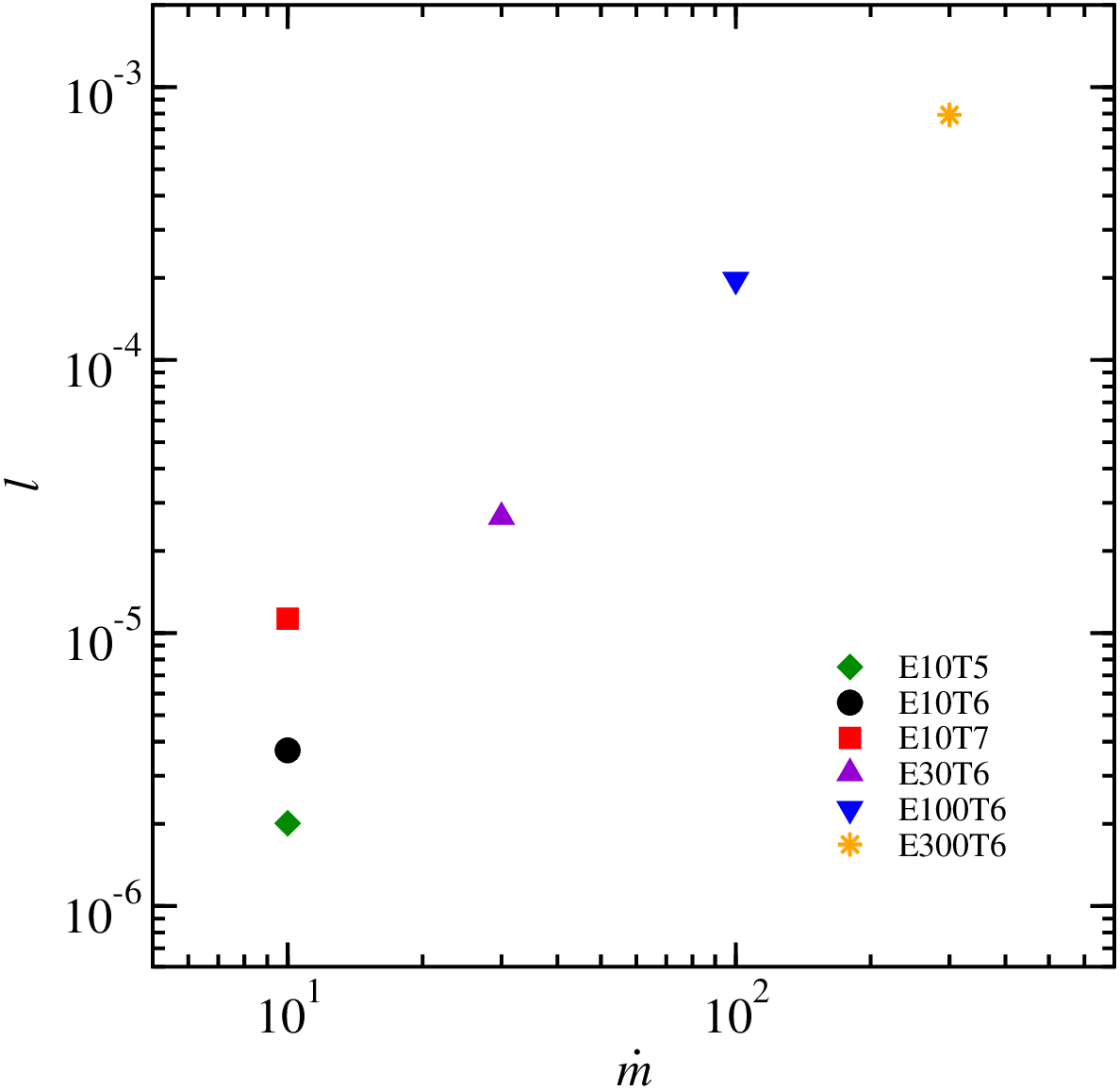}
\caption{Luminosity $l \equiv L/L_\mathrm{Edd}$ as a function of mass accretion rate $\dot{m} \equiv \dot{M}/\dot{M}_\mathrm{Edd}$ for our Bondi simulations.
\label{fig:lmdot}}
\end{figure}

Even though the mass accretion rates in all these simulations are highly super-Eddington ($\dot{m} \ge 10$), the luminosities are not ($l \le 7.9\times10^{-4}$) (see Table \ref{tab:thin}).  Clearly the low radiative efficiency of this flow $\eta \equiv l/\dot{m} \le 2.6\times10^{-6}$ indicates that not all of the binding energy that is liberated by accretion is able to escape in the form of radiation.  Instead, much of the energy, in the form of kinetic energy, heat, and radiation, is advected into the black hole.  Figure \ref{fig:lmdot} presents all of our Bondi results in the $l$-$\dot{m}$ plane.  These results are broadly consistent with earlier studies of optically thick spherical accretion  \citep[e.g.][]{vitello84,nobili91,zampieri96}, given that different assumptions were used in each work.

\section{Conclusion}
\label{sec:conclusion}

In the study of black hole accretion, it only makes sense to treat the whole problem (gravitation, gas dynamics, magnetic fields, and photons) in a relativistic framework.  Therefore, the development of a fully relativistic radiation MHD numerical code is an essential step.  Along with allowing us to answer some of the very longstanding questions concerning black hole accretion, having this capability will also allow us to explore such novel effects as disk self-illumination due to light bending near the black hole.  There are also astrophysical applications of a relativistic radiation MHD code beyond black hole accretion, including core-collapse supernovae, collapsars, and gamma ray burst sources, meaning this work will potentially open up new avenues of research.

This paper demonstrates that we have taken the first step toward that goal.  We now have a scheme that is able to self-consistently treat the radiation on equal footing with the MHD in a fully general relativistic framework.  We have shown that the method performs well when restricted to the appropriate parameter space, i.e. we require $\vert F \vert \equiv \sqrt{F_\mu F^\mu} \ll E$, $\tau \gtrsim 1$, and that the gas and radiation be far from local thermodynamic equilibrium.

One improvement to our method might be to extend it to treat both optically thick and thin flows by implementing a more general closure relation.  Currently, we assume the radiation pressure has the form ${\cal P}^{\hat{i} \hat{j}} = {\cal P} \delta^{\hat{i} \hat{j}} = E/3 \delta^{\hat{i} \hat{j}}$, i.e. we adopt an Eddington factor of $1/3$ to close the set of radiation moment equations.  Although the simplicity of this prescription is desirable, this approximation is only appropriate in the optically thick limit.  More general Eddington tensors, that recover the correct asymptotic behavior for both optically thick and optically thin gas, are available \citep[e.g.][]{levermore84} and could be implemented.

A related issue that must be addressed is that HRSC schemes fail to treat the optically thick limit properly when the photon mean-free path is shorter than the numerical grid spacing \citep{lowrie01}.  The simplest fix for this is to phase out the relevant terms in the diffusion limit when the mean-free path is short \citep{jin96}. 

Finally (and not surprisingly), we have found that the fully explicit method described in this paper suffers from severe timestep restrictions.  Even the relatively modest 512 zone, 1D spherical simulations in \S \ref{sec:bondi}, required approximately 10 million cycles to complete, which took approximately 48 hours on two, dual-core 2.0 GHz AMD Opteron processor.  Clearly a well-resolved 3D disk simulation is out of the question.  One way we could get around this would be to develop a hybrid explicit-implicit scheme, where an implicit step is used to either solve the radiation source terms (while using the current HRSC method to handle the transport) \citep[c.f.][]{turner01} or the full radiation equations (while still solving the MHD equations explicitly).  Unfortunately, implementing implicit solvers in large, multidimensional simulations can be computationally challenging, as it involves the inversion of a large, sparsely populated matrix.  It will take some work to determine the best way to proceed.

\acknowledgements
We thank Brian Farris, Olindo Zanotti, Eirik Endeve, and Scott Noble for their helpful feedback and discussions.  We also thank the anonymous referee.  This work was supported in part by a High-Performance Computing grant from Oak Ridge Associated Universities/Oak Ridge National Laboratory and by the National Science Foundation under Grant No. NSF PHY11-25915.  This research used resources of the Oak Ridge Leadership Computing Facility, located in the National Center for Computational Sciences at Oak Ridge National Laboratory, which is supported by the Office of Science of the Department of Energy under Contract DE-AC05-00OR22725.  MR gratefully acknowledges the support of Summer Undergraduate Research with Faculty (SURF) grant from the College of Charleston.

\clearpage

\end{document}